\documentclass[dvipdft]{article}
\usepackage{graphicx}
\newcommand{\doublespacing}{\let\CS=\@currsize\renewcommand{\baselinesstrech}
{2.0}\tiny\CS}
\begin{document}
\textwidth 16cm
\newcommand{\bd}{\begin{document}}
\newcommand{\ed}{\end{document}}
\newcommand{\bc}{\begin{center}}
\newcommand{\ec}{\end{center}}
\newcommand{\bfr}{\begin{flushright}}
\newcommand{\efr}{\end{flushright}}
\newcommand{\lt}{\left}
\newcommand{\rt}{\right}
\newcommand{\vs}{\vspace}
\newcommand{\hs}{\hspace}
\newcommand{\beq}{\begin{equation}}
\newcommand{\eeq}{\end{equation}}
\newcommand{\lb}{\linebreak}
\newcommand{\pb}{\pagebreak}
\newcommand{\mb}{\makebox}
\newcommand{\fb}{\framebox}
\newcommand{\mc}{\multicolumn}
\newcommand{\ben}{\begin{enumerate}}
\newcommand{\een}{\end{enumerate}}
\newcommand{\bit}{\begin{itemize}}
\newcommand{\eit}{\end{itemize}}
\newcommand{\ol}{\overline}
\newcommand{\un}{\underline}
\newcommand{\lefq}{\lefteqn}
\newcommand{\ba}{\begin{array}}
\newcommand{\ea}{\end{array}}
\newcommand{\beqa}{\begin{eqnarray}}
\newcommand{\eeqa}{\end{eqnarray}}
\newcommand{\beqas}{\begin{eqnarray*}}
\newcommand{\eeqas}{\end{eqnarray*}}
\newcommand{\bfg}{\begin{figure}}
\newcommand{\efg}{\end{figure}}
\newcommand{\bds}{\begin{displaymath}}
\newcommand{\eds}{\end{displaymath}}
\newcommand{\btb}{\begin{tabbing}}
\newcommand{\etb}{\end{tabbing}}
\newcommand{\para}{\parallel}
\newcommand{\pad}{\partial}
\newcommand{\nn}{\nonumber}
\newcommand{\la}{\leftarrow}
\newcommand{\ra}{\rightarrow}
\newcommand{\lgla}{\longleftarrow}
\newcommand{\lgra}{\longrightarrow}
\newcommand{\La}{\Leftarrow}\newcommand{\Ra}{\Rightarrow}
\newcommand{\Lra}{\Leftrightarrow}
\newcommand{\Lgla}{\Longleftarrow}
\newcommand{\Lgra}{\Longrightarrow}
\newcommand{\bm}{\boldmath}
\newcommand{\lan}{\langle}
\newcommand{\ran}{\rangle}
\renewcommand{\a}{\alpha}
\renewcommand{\b}{\beta}
\newcommand{\g}{\gamma}
\newcommand{\G}{\Gamma}
\renewcommand{\d}{\delta}
\newcommand{\eps}{\epsilon}
\newcommand{\Th}{\Theta}
\newcommand{\s}{\sigma}
\newcommand{\lam}{\lambda}
\newcommand{\D}{\Delta}
\newcommand{\vare}{\varepsilon}
\newcommand{\pr}{\prime}
\newcommand{\ro}{\rho}
\newcommand{\nab}{\nabla}
\newcommand{\m}{\mu}
\newcommand{\n}{\nu}
\newcommand{\Sg}{\Sigma}
\newcommand{\p}{\pi}
\newcommand{\R}{I\!\!R}
\newcommand{\om}{\omega}
\newcommand{\Om}{\Omega}
\newcommand{\ze}{\zeta}
\newcommand{\vart}{\vartheta}
\newcommand{\tri}{\triangle}
\newcommand{\f}{\frac}
\newcommand{\iny}{\infty}
\newcommand{\pro}{\propto}

\title{${\cal PT}$ symmetry of a conditionally exactly solvable 
potential}

\author{A.Sinha\thanks{E-mail : anjana23@rediffmail.com}\\ 
Department of Applied Mathematics \\ 
Calcutta University \\ 92 APC Road, 
Kolkata - 700 009, India \\ \\
G. L\'evai\thanks{E-mail : levai@atomki.hu}\\
Institute of Nuclear Research of the Hungarian Academy of Sciences \\ 
P.O. Box 51, H-4001 Debrecen, Hungary\\ \\
and \\ \\ P.Roy\thanks{E-mail : pinaki@isical.ac.in}
\\ Physics and Applied Mathematics Unit,\\ 
Indian Statistical Institute \\ Kolkata  700 108, India}

\maketitle

\begin{abstract}
A conditionally exactly solvable potential, the supersymmetric partner 
of the harmonic oscillator is investigated in the ${\cal PT}$-symmetric
setting. It is shown that a number of properties characterizing
shape-invariant and Natanzon-class potentials generated by an imaginary
coordinate shift $x-{\rm i}\epsilon$ also hold for this potential 
outside the Natanzon class. 
\end{abstract}




\section{Introduction}

The concept of ${\cal PT}$ symmetry has generated much interest recently 
in one-dimensional quantum mechanical potential problems 
as many problems exhibiting ${\cal PT}$ symmetry, i.e. 
invariance under the simultaneous action of space (${\cal P}$) and time 
(${\cal T}$) inversion possessed real energy eigenvalues belonging to 
the discrete spectrum, although the correspondig Hamiltonians were not
Hermitian \cite{ben1}. Later it was found that in contrast with the first 
conjectures, ${\cal PT}$ symmetry is neither necessary, nor sufficient 
condition for having real discrete spectrum.
More recently ${\cal PT}$ symmetry 
was recognized as a special case of $\eta$-pseudo-Hermiticity 
\cite{mosta}: a Hamiltonian is $\eta$-pseudo-Hermitian 
if there exists a linear, Hermitian, invertible operator $\eta$, for which 
$H^{\dagger} = \eta H \eta^{-1}$ holds. 
In this context ${\cal PT}$ symmetry is ${\cal P}$-pseudo-Hermiticity for 
one-dimensional Hamiltonians of the type $H=p^2+V(x)$, 
whereas conventional Hermiticity 
follows for $\eta=1$. More recently the formalism of ${\cal 
PT}$ symmetry has been interpreted as the complex extension
of quantum mechanics, by modifying it with the help of a dynamically 
constructed ${\cal C}$ operator, so that the inner product 
$\langle \psi_i | {\cal CP} | \psi_j \rangle$ 
leads to positive norm \cite{ben3}. 

On the other hand, exactly solvable problems are 
of enormous importance in the understanding of physical systems, 
and this is also the case in ${\cal PT}$-symmetric quantum mechanics. 
The well-known textbook examples (e.g. the harmonic oscillator, Coulomb, 
Morse, P\"oschl--Teller, Rosen--Morse, etc.) potentials belong to a rather 
narrow two- and three-parameter subset of the general six-parameter 
Natanzon potentials \cite{jpa03-25}. In particular, they are 
shape-invariant \cite{jpa03-13,jpa02-27} potentials 
having the property that a supersymmetric transformation 
eliminating their ground state does not modify the functional form of the 
potential, but only changes some parameters appearing in them. Supersymmetric 
quantum mechanics (SUSYQM) \cite{jpa02-3} has been a rather productive 
method of generating new exactly solvable potentials from known ones. The 
SUSY partner potentials generated this way have the same discrete energy 
spectrum, except perhaps a single level which is eliminated (the ground 
state), or is added to the spectrum (below the ground state). Applying a 
SUSYQM transformation to a general 
Natanzon (i.e. non-shape-invariant) 
potential results in a SUSY partner potential that is outside the Natanzon 
class, because in this case the bound-state wavefunctions can be written 
in terms of two (confluent) hypergeometric functions.

To widen the class of exactly solvable models of the Schr\"{o}dinger
equation, another concept used is 
conditional exact solvability (CES), rendering the potential exactly 
solvable only when the potential parameters appearing in them 
satisfy certain conditions. 
For example, the DKV potential was proven to be a Natanzon-class 
potential \cite{jpa03-31}, and its CES nature stems from the fact that it 
has three potential terms, but only two free parameters. Another type of 
CES potentials has been identified in the SUSYQM construction: in this 
case the supersymmetric partner of shape-invariant potentials was 
constructed by inserting a new ground state below the original one, and it 
was found that this could be done at certain energies, which required 
setting a parameter to a numerical constant \cite{jun}. With this a 
CES potential outside the Natanzon class could be generated.  

Constructing the ${\cal PT}$-symmetric version of exactly solvable 
potentials resulted in a number of interesting findings. It turned out, 
for example, that except for the Coulomb and Morse potentials all the 
shape-invariant potentials can be defined on a trajectory determined by 
the imaginary coordinate shift $ x \rightarrow  x - {\rm i} \epsilon $  
\cite{jpa03-7,jpa03-20}. (The Coulomb \cite{jpa03-16} and Morse 
potentials \cite{jpa03-15} possess normalizable solutions only on 
some curved trajectories of the complex $x$ plane.)  
With the appropriate choice of the parameters, 
normalizable solutions with both real and complex energies could be 
generated in a straightforward way for all the remaining shape-invariant 
potentials \cite{jpa03-7,jpa03-20}. It also turned out that with the 
imaginary coordinate shift 
the singularities of real potentials (e.g. at the origin) could be 
cancelled, and as the result of this, the radial potentials could be 
extended to the whole $x$ axis formally, and new normalizable solutions 
appeared due to the less strict boundary conditions. Moreover, 
these potentials had {\it two} series of normalizable states, 
distinguished by the $q=\pm 1$ quasi-parity quantum number, which 
characterizes the bound states, but the potential itself does not depend 
on it \cite{jpa03-19}. This gave rise to 
{\it two} different (`fermionic') SUSY partners due to the presence of 
{\it two} nodeless solutions (with quasi-parity $q=1$ and $q=-1$) 
to the original (`bosonic') potential \cite{jpa02-34,jpa02}.
Furthermore, it was also proven that in case the 
original potential has unbroken ${\cal PT}$ symmetry, then the two partner 
potentials also have this property, but if the ${\cal PT}$ symmetry 
of the original potential is spontaneously broken, then the partner 
potentials cease to be ${\cal PT}$-symmetric. This finding proved valid 
for some shape-invariant potentials (e.g. the Scarf II potential 
\cite{jpa02}) and also for some Natanzon-class potentials (e.g. the 
generalized Ginocchio potential) \cite{gin}. 

All these results raise a number of questions concerning the properties 
under ${\cal PT}$ symmetry of various types of solvable potentials. A 
number of results seem to indicate that properties characterizing mainly 
${\cal PT}$-symmetric shape-invariant potentials are actually, valid for 
more general potentials from the Natanzon class too. This is the case with 
the applicability of the imaginary coordinate shift, the presence of the 
quasi-parity quantum number and the behaviour of the SUSY partner 
potentials in the case of intact and spontaneously broken ${\cal PT}$ 
symmetry of the original potential. It is natural to ask whether these 
properties also characterize potentials even beyond the Natanzon class. A 
natural candidate for these studies is the SUSY partner of shape-invariant 
potentials generated by inserting a new ground state below the original 
one. The CES potential generated in this way is outside the Natanzon 
class, as we have discussed before \cite{jun}. However, all its main 
characteristics can be determined in terms of exact calculations, so we 
can hope to get answers to our questions formulated here.

\section{A conditionally exactly solvable ${\cal{PT}}$-invariant 
potential}

In this section, our aim is to construct CES potentials which are SUSY
partners of the much studied ${\cal{PT}}$-symmetric harmonic oscillator 
\cite{jpa03-6}
\beq
V(x) = (x - i \eps) ^2 + \f{ \alpha ^2 - \f{1}{4} }{ (x - i \eps )^2}\ .
\label{pto} 
\eeq
Shifting the co-ordinate from $x$ to $ z ~=~ x - i \eps $, removes the
singularities on the real line, and extends the potential from the half line 
to the full line. The eigenvalues and eigenfunctions of (\ref{pto}) are well
known \cite{jpa03-6}
\beq
\psi _{nq} = N_{nq} e^{- z ^2 / 2 }  
z ^{-q \a + 1/2} L_n ^{(-q \a)} \lt( z ^2 \rt) \ ,
\eeq
\beq
E_{nq} = 4n +2 -2q \a \ ,
\eeq
where $L_n ^{(\sigma)} (z^2) $ are the associated 
Laguerre polynomials \cite{handbook}
and $q = \pm 1$ is the quasi-parity \cite{jpa03-19}.

We wish to find Hamiltonians which are isospectral to (\ref{pto}), with the
possible exception of the ground state. For this purpose
we define two intertwining operators $A_{(q)}$ and $B_{(q)}$ 
\beq
A_{(q)} = \f{\rm d}{{\rm d}x} + W_q (x)
\eeq
and
\beq
B_{(q)} = - \f{\rm d}{{\rm d}x} + W_q (x)\ ,
\eeq
where $W_q (x)$ (the so-called superpotential in conventional Hermitian 
quantum mechanics), is, in general, a complex-valued function. It is 
easy to observe that if $ \psi ^+ _q (x)$ 
is an eigenfunction of $H_+ ^{(q)}$ with eigenvalue $E_q$, then 
$ \psi ^- _q (x) = B \psi ^+ _q (x) $ is an eigenfunction of  
$H_- ^{(q)}$ with the same eigenvalue $E_q$, 
where the partner Hamiltonians $ H_{\pm}$ are given by
\beq
H_{+} ^{(q)} = A_{(q)} B_{(q)} - \beta _q 
= -\f{{\rm d}^2}{{\rm d}x^2} + V_{+}^{(q)} (x) - \beta _q
\eeq 
\beq
H_{-} ^{(q)} = B_{(q)} A_{(q)} - \beta _q 
= -\f{{\rm d}^2}{{\rm d}x^2} + V_{-}^{(q)} (x) - \beta _q
\eeq
with 
\beq
V_{\pm} ^{(q)} (x) = W_{(q)} ^2 (x) \pm W_{(q)} ^{\pr} (x) \ .
\eeq
It is worth mentioning here that $ \beta _q $ can be $q$-dependent. In fact,
as we shall see later, $ \beta _q $ turns out to be the $q$-dependent
factorization energy. 
The interesting point to note is that unlike in conventional
Hermitian quantum mechanics, $A_q$ and $B_q$ are not
mutually adjoint operators. They may be related by a linear, invertible, 
Hermitian operator $\eta$ to form mutually pseudo-adjoint pair
\cite{mostafa2}.  
The role of the quasi-parity quantum number $(q = \pm 1)$, is quite
important as it gives rise to a doublet set of isospectral partners for the
original potential.

To construct  isospectral, non-shape-invariant partners of the
${\cal{PT}}$-symmetric oscillator, 
we assume the following ansatz for $W_{(q)} (x)$ :
\beq
W_{(q)} (x) = (x-i \eps) + \f{\lam}{(x-i \eps)} + \sum ^N _{k=1}
\f{ 2 g_k (x-i \eps)}{1 + g_k (x-i \eps) ^2 } \ \ \ \ \ , 
\ \ \ \ \ g_k \geq 0\ ,
\eeq
which reduces to the superpotential of the ${\cal{PT}}$-symmetric oscillator
for \\
$ g_1 ~=~ g_2 ~=~ g_3 ~=~ \cdots ~=~ g_N ~=~ 0 $. Thus the supersymmetric
techniques applied here are different from the standard ways when the ground
state is eliminated. Rather this is a reverse procedure in a sense, in which
$V_+$ is a simple potential and we construct $V_-$ which has one more state. 
Further, in this work we restrict
ourselves to the case $N=1$. The partner potentials then assume the form
\beq
V^{(q)}_+ (x) ~=~ (x - i \eps)^2 ~+~ 
\f{\lam \lt( \lam -1 \rt)}{(x - i \eps)^2} ~+~
\f{4g \lam + 2 g -4}{1 + g \lt( x - i \eps \rt) ^2} ~+~ 2 \lam ~+~ 5\ ,
\eeq
\beq
V^{(q)}_- (x) ~=~ (x - i \eps)^2 
~+~ \f{\lam \lt( \lam +1 \rt) }{(x - i \eps)^2} 
~-~ \f{4 g \lam - 2 g - 4}{1 + g \lt( x - i \eps \rt )^2} 
~+~ \f{8 g ^2 (x - i \eps )^2}{\lt\{1  + g (x-i \eps )^2 \rt \} ^2 } 
~+~ 2 \lam ~+~ 3\ .
\eeq 
Since the functional form of $V^{(q)}_+(x)$ should be of the same form 
as that of $V(x)$ in (\ref{pto}), simple algebra shows that 
\beq
\lam = -q \alpha + \f{1}{2}\ .
\eeq
This is an example of CES problem, as exact solvability occurs only when the
potential parameter $g$ assumes the specific value 
\beq
g = \f{1}{ -q \alpha +1 } \label{g}
\eeq
reducing the partners to 
\beq
\ba{lcl}
\displaystyle v_+ (x) 
&\equiv& \displaystyle V^{(q)}_+ (x) - \beta _q \\
&=& \displaystyle (x - i \eps)^2 
+ \f{\alpha ^2 -\f{1}{4}}{(x - i \eps)^2} + 6\ ,
\ea
\eeq
\beq
\ba{lcl}
\displaystyle v^{(q)}_- (x) 
&\equiv& \displaystyle V^{(q)}_- (x) - \beta _q \\
&=& \displaystyle (x - i \eps)^2 
+ \f{\alpha ^2 - 2q \alpha + \f{3}{4} }{(x - i \eps)^2} 
- \f{4}{1-q \alpha + (x - i \eps)^2} 
+ \f{8(x - i \eps )^2}{\lt[ (-q \alpha +1)  
+  (x-i \eps )^2 \rt] ^2 } + 4\ .
\label{partner}
\ea
\eeq 
Note that after the substitution of $g$ from (\ref{g}) $v_+(x)$ is 
independent from $q$, while $v^{(q)}_-(x)$ is not. Equation (\ref{partner}) 
is an example of a CES potential, for the particular
value of $g$ given in (\ref{g}). Moreover, it has more terms than its partner
$v_+ (x)$, and hence cannot have the same functional form. Thus we obtain a 
non-shape-invariant isospectral partner of the ${\cal{PT}}$-symmetric
oscillator. 
In the above $ \beta _q $ stands for 
\beq
\beta _q = - 2q \alpha\ .
\eeq
Thus to each $ v_+ (x) $, 
there exist two non-shape-invariant, isospectral partners 
$ v^{(q)}_-(x)$, as shown above. The interesting feature observed here is that there
are no singularities on the real line, and hence the CES potentials so
constructed are defined on the full line $( - \infty, + \infty )$. \\
The possible zero-energy eigenfunctions of 
$H_{\pm}^{(q)}+\beta _q$ are of the form
\beq
\psi _{0q} ^{(\pm)} (x) = N_{0q} 
~ \exp \lt( \pm \int ^x ~W_{(q)}(t) {\rm d}t \rt)\ ,
\eeq
where $N_q$ is the normalization constant. Since $ \psi _{0q} ^{(-)} (x) $
given by
\beq
\displaystyle \psi _{0q} ^{-} (x) ~=~ 
N_{0q} \f{1}{ 1 + g (x-i \eps ) ^2} 
e^{ - \f{ (x-i \eps )^2}{2}} (x - i \eps) ^{- q \a +\f{1}{2}}
\eeq
is normalizable, the situation may be compared to that of unbroken
supersymmetry. So the eigenfunctions $ \psi _{nq} ^{(\pm)} $ 
and eigenvalues $ E_{nq} $ of the partner Hamiltonians 
$ H_{\pm} ^{(q)}$ are related by ($n = 0, 1, 2, \dots$)
\beq
E_{0q} ^{-} = 0, \ \ \ \ E_{(n+1)q} ^{-} = E_{nq} ^{+} > 0 \ ,
\eeq
\beq
\psi _{nq} ^{+} = (E_{(n+1)q} ^{-})^{-1/2} A_{(q)} 
 \psi _{(n+1)q} ^{(-)}\ ,
\eeq
\beq
\psi _{(n+1)q} ^{-} = (E_{nq} ^{+})^{-1/2} B_{(q)} 
\psi _{nq} ^{(+)}\ .
\eeq
Since the eigenfunctions and eigenvalues for the Hamiltonian  
$H_{(q)} ^{+}$ are well known, 
the wave functions of the partner Hamiltonian 
$H_{(q)} ^{-} $ are calculated to be
\beq
\ba {lcl}
\displaystyle \psi _{(n+1)q} ^{-} 
&=& N_{nq}(E_{(n+1)q} ^{+})^{-1/2}
 ~ \exp \lt( -\f{z^2}{2} \rt)~ z^{ -q \alpha + \f{1}{2} } \\
&\times& \left[ \lt( 2z + \f{2gz}{1 + g z^2} -\f{2(n+1)}{z} \rt) 
 L_{n+1} ^{(-q \alpha)} \lt( z^2 \rt) ~-~ \f{2(n - q \alpha +1)}{z}
~ L_{n} ^{(-q \alpha)} \lt( z^2 \rt) \right]\ . 
\label{wavepart}
\ea
\eeq
From the structure of (\ref{partner}) and (\ref{wavepart})
it is evident that $v^{(q)}_-(x)$ belongs to a class of
potentials which is beyond the shape-invariant as well as the Natanzon class
potentials. However, in analogy with the shape-invariant and Natanzon potentials,
the states are characterized by the quasi-parity $q$, giving rise to two SUSY
isospectral partners  for the original potential. 

Now let us analyse the conditions for having real and complex energy
eigenvalues. As we shall see the role played by 
the potential parameter $ \a $ is very crucial in this
regard.

\vs{1cm}

\section{${\cal{PT}}$ symmetry of the SUSY partner}

\noindent
(i) ~~~ $ \alpha $ is real : 
The ${\cal{PT}}$ symmetry of the original potential $v_{+}(x)$ is 
unbroken. 

The parameter $g$ turns out to be real in this case, and
the two partner potentials $v^{(q)}_{-} (x)$ (with $q=\pm 1$) are also 
${\cal{PT}}$-invariant, as can be seen from the behaviour of their real 
and imaginary components, which are even and odd functions of $x$,
respectively. Figure 1 shows the imaginary parts of the original potential 
together with its two partners (corresponding to $q=\pm 1$) for $\alpha=0.3$ and $\epsilon=0.5$ while figure 2 shows the real parts of the same potentials for identical parameter values.

\vs{.5cm}

\noindent
(ii) ~~~ $ \alpha $ is pure imaginary : 
${\cal{PT}}$ symmetry spontaneously broken in the original potential 
$v_{+}(x)$. 

Let $ \alpha = i a $, where $a$ is real. 
This choice of $ \alpha $ renders $g$ to be complex. The 
${\cal{PT}}$-invariant $v_{+}(x)$ sector has an attractive, but 
non-singular core
\beq
v_+ (x) = (x - i \eps)^2 - \f{a ^2 +\f{1}{4}}{(x - i \eps)^2} + 6
\eeq
with complex conjugate pairs of energies
\beq
E_{nq} ^{+} = 4n +8 - i2qa\ .
\eeq
Though there still exist two values of $g$ and consequently of 
$v^{(q)}_-(x)$, the partners given by 
\beq
v^{(q)}_- (x) = (x - i \eps)^2 
+ \f{-a ^2 - 2iqa + \f{3}{4} }{(x - i \eps)^2} 
- \f{4}{1-iqa + (x - i \eps)^2} 
+ \f{8(x - i \eps )^2}{\lt[ 1-iqa  +  (x-i \eps )^2 \rt] ^2 } + 4 
\eeq 
are no longer ${\cal{PT}}$-invariant. 
Figure 3 shows the imaginary parts of the original potential 
together with its two partners (corresponding to $q=\pm 1$) for $\alpha=0.3i$ and $\epsilon=0.5$ while figure 4 shows the real parts of the same potentials for identical parameter values. 

As illustrative examples 
we calculate the ground state and the first excited state 
wave functions for this case:

\beq
\psi _{0q} ^{-} = C_{0q} (1 - iqa) ~ e ^{ -\f{z^2}{2}} ~ 
\f{z^{ -iqa + \f{1}{2} } }{1 -iqa + z^2} \ ,
\eeq
\beq
\psi _{1q} ^{-} = C_{1q}  ~ e ^ { -\f{z^2}{2} }
~ z^{ -iqa + \f{1}{2} } 
~ \lt[ \lt( 2z + \f{2gz}{1 + g z^2} -\f{2}{z} \rt)
~ \lt( 1 - iqa - z^2 \rt) ~-~ \f{2(1 - iqa )}{z} \rt]\ . 
\eeq
where $C_{0q}$ and $ C_{1q}$ are some normalization factors. 
It is clearly seen that the eigenfunctions are no longer 
${\cal{PT}}$-invariant, rather the ${\cal PT}$ operation 
transforms $\psi_{nq_+}$ and $\psi^-_{nq_-}$ ($q_{\pm}=\pm 1$) into each other. It is worthwhile to note that this relation also holds between $v^{(q)}_-$ and $v^{(-q)}_-$ in the case of imaginary $\alpha$.

\section{Summary and conclusions}

We analysed a conditionally exactly solvable potential, the 
supersymmetric partner of the ${\cal PT}$-symmetric harmonic oscillator, 
which is outside the Natanzon potential class by construction. Our 
motivation was to investigate whether some typical features originally
found for the ${\cal PT}$-symmetric version of most shape-invariant 
potentials and later proved also for some non-shape-invariant 
Natanzon-class potentials (the generalized Ginocchio potential) 
generated by an imaginary
coordinate shift $x-{\rm i}\epsilon$ remain valid for this kind of
potential too. These features included the presence of the quasi-parity
quantum number $q=\pm 1$, the ``sudden'' realization of the spontaneous
breakdown of ${\cal PT}$ symmetry (i.e. the simultaneous disappearance
of real energy eigenvalues and their re-emergence as complex conjugated
pairs at a certain value of a parameter) and the finding that the
spontaneous breakdown of the original potential implies the manifest
breakdown of the ${\cal PT}$ symmetry of its two supersymmetric partner.
Our study confirmed that all these features are valid in this case too,
so they characterize a much wider potential class than originally
thought. It seems that these features appear for all the 
${\cal PT}$-symmetric potentials that are generated by an imaginary
coordinate shift. Certainly they are absent in shape-invariant 
\cite{jpa03-16,jpa03-15} and non-shape-invariant Natanzon-class
\cite{pla01} potentials defined on curved trajectories of the complex 
$x$ plane. Further work is needed for the detailed analysis of 
these differences.

\section*{Acknowledgment}

One of the authors (A.S.) acknowledges financial assistance from CSIR, India.
This work was supported by the OTKA grant No. T031945 (Hungary) and by 
the MTA--INSA (Hungarian--Indian) cooperation.

\pb
\center{Figure Captions}

\begin{enumerate}
\item[] Fig 1. Imaginary parts of the original potential (solid line), the partners (dashed line for $q=1$, dotted line for $q=-1$) for unbroken $\cal{PT}$ symmetry for $\alpha=0.3, \eps=0.5$.

\vspace{0.2cm}

\item[] Fig 2. Real parts of the original potential (solid line), the partners (dashed line for $q=1$, dotted line for $q=-1$) for unbroken $\cal{PT}$ symmetry for $\alpha=0.3, \eps=0.5$.

\vspace{.2cm}

\item[]Fig 3. Imaginary parts of the original potential (solid line), the partners (dashed line for $q=1$, dotted line for $q=-1$) for spontaneously broken $\cal{PT}$ symmetry for $\alpha=0.3i,\eps=0.5$.

\vspace{.2cm}

\item[] Fig 4. Real parts of the original potential (solid line), the partners (dashed line for $q=1$, dotted line for $q=-1$) for unbroken $\cal{PT}$ symmetry for $\alpha=0.3i, \eps=0.5$.
\end{enumerate}

\pb

\ed